%% file: QuantumPiggyR1.tex
\documentclass[aps,pra,reprint,groupedaddress]{revtex4-1}
\usepackage{amsmath}
\usepackage{amsfonts}
\usepackage{amssymb}
\usepackage{amsthm}

\usepackage{blindtext}
\usepackage{graphicx}
\usepackage{dcolumn}
\usepackage{bm}

\usepackage{listings}
\usepackage{matlab-prettifier} 

\usepackage{framed}         
\usepackage{soul}           
\usepackage{lmodern} 

\usepackage{color} 

\usepackage{rotating}
\usepackage{floatpag}
\rotfloatpagestyle{empty}

\usepackage{graphicx}

\usepackage{mathrsfs} 
\usepackage{longtable}
\usepackage{physics}

\usepackage{mathtools,bm}

\usepackage{cancel}

\usepackage[nolist]{acronym} 

\usepackage[yyyymmdd,hhmmss]{datetime} 
\newdateformat{monthyeardate}{\monthname[\THEMONTH] \THEDAY, \THEYEAR} 
\usepackage[pagewise,mathlines,displaymath]{lineno} 

\input{WinsNotation.sty}
\input{PiggyNotation.sty}

\begin{document}

\title{Piggybacking on Quantum Streams}

\author{Marco Chiani} \email{marco.chiani@unibo.it}\affiliation{University of Bologna, Italy}
\author{Andrea Conti}\email{a.conti@ieee.org}\affiliation{University of Ferrara, Italy}
\author{Moe Z. Win}\email{moewin@mit.edu}\affiliation{Massachusetts Institute of Technology, USA}

\begin{abstract}
This paper shows that it is possible to piggyback classical information on a stream of qubits protected by quantum error correcting codes.
The piggyback channel can be created by introducing intentional errors corresponding to a controlled sequence of syndromes. These syndromes are further protected, when quantum noise is present, by \aclp{CECC} according to a performance--delay trade-off.
Classical information can thus be added and extracted at arbitrary epochs without consuming additional quantum resources and without disturbing the quantum stream. 
\end{abstract}
\maketitle
\input{acronymsMC.tex}  

\section{Introduction}
We foresee the possibility of piggybacking classical information on a stream of qubits protected by a \ac{QECC}. To this end, we propose a method to send a
sequence of classical bits on quantum streams by introducing intentional noise. 
This  noise  
induces a controlled sequence of syndromes, which can be measured without destroying quantum superposition.  
The syndromes 
can then be used to encode classical information on top of quantum streams, enabling several possible applications. 
In particular, piggybacking on quantum streams can facilitate control and annotation 
for quantum systems and networks.

Consider for example a network in which nodes exchange quantum information among each other  \cite{Zol:05,kim:08,Van:13,DaiPenWin:J20b,Weh:18,NAP:19,OSW18}. 
In addition to user data, control data such as synchronization patterns, node addresses, and routing parameters are needed for the network operation. 
In classical networks, control data consume physical resources. 
For instance, in-band synchronization requires that transmitting nodes insert specific patterns of bits into data streams (consuming additional bandwidth) to delimit packets, and that the receiving nodes search for such patterns from incoming bits \cite{SuwChiWin:J08}. 
However, inserting qubits as control data is not a viable option for quantum networks, since measuring can destroy quantum state superposition 
 \cite{NieChu:B10}.  
For this reason, several studies assert that quantum networks will need classical networks for out-of-band signaling and control \cite{OSW18}. 
On the other hand, transmission of classical bits together with random numbers for \ac{QKD} (using continuous variables) was developed in
\cite{Qi:16,WuWanLiaZhoGuo:19,KumWonPenSpiWhi:19} for security enhancement of classical networks. Instead, we aspire to the transmission of classical bits together with quantum bits (using discrete variables) for control of quantum networks.

A possible way to create a control channel in quantum networks is to introduce dedicated auxiliary orthogonal states.  
For instance, consider a quantum system with qutrits (Hilbert space of dimension three) instead of qubits, where the orthogonal states $\ket{0}, \ket{1}$ are used as the basis for information, and an additional orthogonal state $\ket{2}$ is used for synchonization \cite{Fuj:13}.  Inserting patterns of states $\ket{2}$ in different positions along the quantum stream can also carry classical information. 
Note that systems employing qutrits have read-only capability  
since altering the classical information would require changing the positions of the $\ket{2}$'s.     
Besides this limitation, the main difficulty is the need for working with qutrits instead of the usual qubits, thereby impacting the overall system architecture. 

In this work, we propose a new method to write, read, and eventually rewrite classical information by piggybacking on top of quantum streams.  
Our method can add and extract the classical information at arbitrary epochs without consuming additional quantum resources and without disturbing the quantum stream.  
This technique enables new capabilities by unleashing a hidden classical channel provided by \acp{QECC}. 

\vspace{-0.2cm}
\section{Piggybacking via intentional errors}
This section first introduces the notation and the main elements of \acp{QECC}, then it proposes the idea of piggybacking classical information on a quantum stream for noiseless and noisy quantum channels.
\subsection{Preliminaries}
Consider quantum information carried by qubits, which are elements of the two-dimensional Hilbert space $\mathcal{H}^{2}$, with basis $\ket{0}$ and $\ket{1}$ \cite{NieChu:B10}. 
An $n$-tuple of qubits ($n$ qubits) is an element of the $2^n$-dimensional Hilbert space, $\mathcal{H}^{2^n}$\!, with basis composed by all possible tensor products $\ket{i_1} \ket{i_2} \cdots \ket{i_n}$, with $i_j \in \{0,1\}, 1\le j\le n$.  
The Pauli operators, denoted as $\M{I}, \M{X}, \M{Z}$, and $\M{Y}$, are defined by  $\M{I}\ket{a}=\ket{a}$, $\M{X}\ket{a}=\ket{a\oplus 1}$, $\M{Z}\ket{a}=(-1)^a\ket{a}$, and $\M{Y}\ket{a}=i(-1)^a\ket{a\oplus 1}$ for $a \in \lbrace0,1\rbrace$. These operators either commute or anticommute. 
The Pauli group $\mathcal{G}_n$ on $n$ qubits is generated by all possible $n-$fold tensor products of these four operators together with the factors $\pm 1$ and $\pm i$.
Two operators in $\mathcal{G}_n$ commute if and only if there is an even number of places where they have different Pauli matrices that are not the identity $\M{I}$.

When necessary ``q-codeword'' and ``c-codeword'' will be used to distinguish 
quantum and classical codewords. In the block diagrams, single lines and double lines  are used for qubits and classical bits, respectively. 
\begin{figure}
	\begin{center}
		\includegraphics[width=1.\columnwidth]{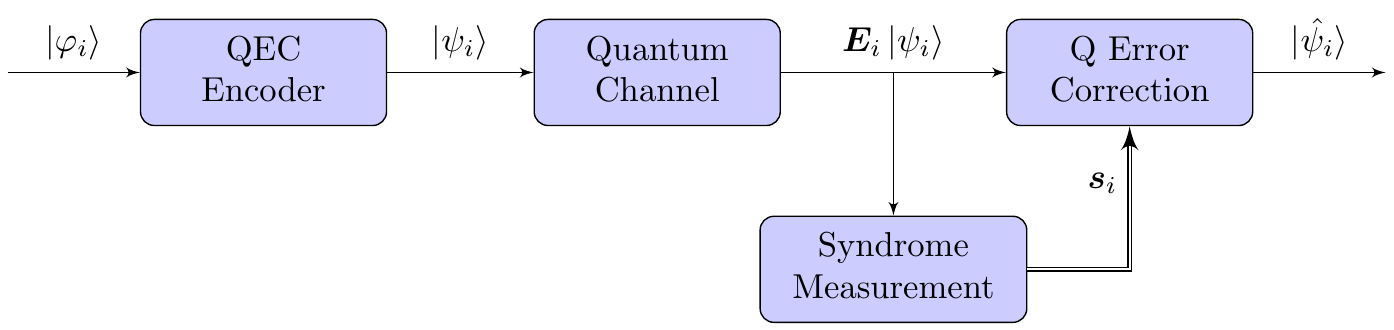}
	\end{center}
	\vspace{-10pt}
	\caption{Quantum communication link employing quantum error correction.}
	\label{fig:nopiggyblockdiagram}
\end{figure}
Fig.~\ref{fig:nopiggyblockdiagram} shows the block diagram of a generic quantum communication link between two nodes employing a \ac{QECC} designed to cope with channel impairments \cite{Sho:95,CalSho:96,Ste:96,LafMiqPazZur:96,Got:96,EkeMac:96,FleShoWin:J07,gottesman2009introduction,BabHan:19}. While the proposed  method is valid for all \acp{QECC}, for the sake of clarity 
our discussion is restricted to the case of block codes.  

Consider an $[[n,k]]$ \ac{QECC} that encodes $k$ data qubits $\ket{\varphi}$ into a codeword of $n$ qubits $\ket{\psi}$.   
Specifically, consider a stabilizer code $\mathcal{C}$ generated by $n-k$ independent and commuting operators $\M{G}_i \in \mathcal{G}_n$, called generators, such that the subgroup generated by these $\M{G}_i$'s does not contain $-\M{I}$   \cite{Got:96,gottesman2009introduction,NieChu:B10}.  
The  code $\mathcal{C}$ is the set of quantum states $\ket{\psi}$ satisfying 
\begin{linenomath}\begin{align}
\label{eq:gj}
\M{G}_i \ket{\psi}=\ket{\psi} \,, \quad  i=1, 2, \ldots, n-k\,.
\end{align}\end{linenomath}

Consider a codeword $\ket{\psi}  \in \mathcal{C}$ affected by a channel error described by the operator $\M{E} \in \mathcal{G}_n$. 
For error correction, the received state $\M{E}\ket{\psi}$ is measured according to the generators $\M{G}_1, \M{G}_2, \ldots, \M{G}_{n-k}$, resulting in a quantum error syndrome $\Syndrome(\M{E})=(s_1, s_2, \ldots,s_{n-k})$, with each $s_i =+1$ or $-1$ depending on the fact that $\M{E}$ commutes or anticommutes with $\M{G}_i$, respectively. Note that, due to \eqref{eq:gj}, the syndrome depends on $\M{E}$ and not on the particular q-codeword $\ket{\psi}$. Moreover, measuring the syndrome does not change the quantum state, which remains $\M{E} \ket{\psi}$ \cite{Got:96,gottesman2009introduction,NieChu:B10}. 
Let $\mathcal{S}=\{\Syndrome^{(1)}, \Syndrome^{(2)}, \ldots, \Syndrome^{(m)}\}$ be the set of $m=2^{n-k}$ possible syndromes, with $\Syndrome^{(1)}=(+1,+1,\ldots,+1)$ denoting the syndrome of the operators $\M{E}$ (including the identity $\M{I}$, i.e., the no-errors operator) such that $\M{E}\ket{\psi}$ is still a valid q-codeword. 

Among the set of channel errors on the $n$ qubits producing the syndrome $\Syndrome^{(i)}$, let $\M{Q}^{(i)}$ denote the operator corresponding to the specific error that can be corrected, and  let $\mathcal{Q}= \{\M{Q}^{(1)}, \M{Q}^{(2)}, \ldots, \M{Q}^{(m)}\}$. 
In other words, if the measured syndrome is $\Syndrome^{(i)}$, the quantum decoder applies 
the recovery operator $\M{Q}^{(i)\dag}$ to produce a valid codeword.  
For example, consider the $[[3,1]]$ repetition \ac{QECC} that can correct one bit-flip error by mapping a qubit $\alpha \ket{0} + \beta \ket{1}$ into a q-codeword $\alpha \ket{000} + \beta \ket{111}$. This code has generators $\M{G}_1= \M{Z} \M{Z} \M{I}$ and $\M{G}_2= \M{I} \M{Z} \M{Z}$, and syndromes $\mathcal{S}=\{(+1,+1), (-1,+1), (-1,-1), (+1,-1) \}$ with corresponding correctable errors $\mathcal{Q}= \{\M{I} \M{I} \M{I} , \M{X} \M{I} \M{I},\M{I} \M{X} \M{I},$ $ \M{I} \M{I} \M{X}\}$. 
 
\subsection{Piggybacking: the basic idea}

The basic idea of piggybacking is described in the following. Consider an $[[n,k]]$ \ac{QECC} that encodes $k$ data qubits $\ket{\varphi}$ into a codeword of $n$ qubits $\ket{\psi}$. 
So, a sequence $\ket{\varphi_1}, \ket{\varphi_2}, \ldots $ of data qubits is encoded into a sequence of q-codewords $\ket{\psi_1}, \ket{\psi_2}, \ldots $. 

The transmitter  inserts intentional errors by employing a sequence of operators $\M{P}_1,\M{P}_2, \ldots$ with corresponding error syndromes $\Syndrome_1, \Syndrome_2, \ldots$ so that the transmitted codewords are $\M{P}_1 \ket{\psi_1},\M{P}_2 \ket{\psi_2}, \ldots$. 
In particular, these intentional errors are chosen from the set of correctable errors, i.e., $\M{P}_i \in \mathcal{Q}$. 
A decoder at the receiver side measures the quantum error syndromes and infers the sequence of intentional errors $\hat{\Syndrome}_1, \hat{\Syndrome}_2, \ldots$.  

\begin{figure}[t]
	\begin{center}
		\includegraphics[width=%
		0.45\columnwidth]{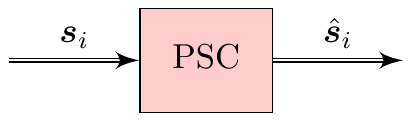}
	\end{center}
	\vspace{-10pt}
	\caption{The \acl{PSC}. 
	}
	\label{fig:piggyblockdiagramPSC}
\end{figure}

The aforementioned procedure creates an $m$-ary discrete-input discrete-output channel with alphabet $\mathcal{S}$ for both input and output symbols. 
This classical channel is  referred to as a \ac{PSC}  (Fig.~\ref{fig:piggyblockdiagramPSC}). 
In the following, we describe in detail the proposed method, first assuming a noiseless quantum channel and then a noisy one.

\subsection{Piggybacking over a noiseless quantum channel}
\begin{figure*}[t]
	\begin{center}
		\includegraphics[width=1.6\columnwidth]{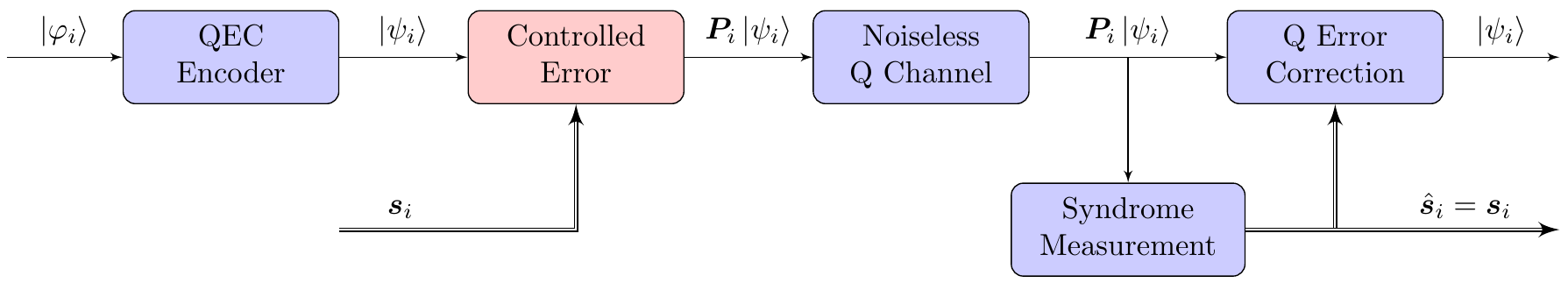}		
	\end{center}
	\vspace{-10pt}
	\caption{Quantum piggybacking by intentional errors, noiseless quantum channel. The block added to realize the piggyback classical channel is in red. 
	}
	\label{fig:piggyblockdiagramNOISELESS}
\end{figure*}

If the quantum channel is noiseless and 
an intentional error is introduced by applying the corresponding operator $\M{P}_i\in \mathcal{Q}$ on the $i$-th q-codeword, the transmitted state will be $\M{P}_i\ket{\psi_i}$ (see  Fig.~\ref{fig:piggyblockdiagramNOISELESS}).   
Since the quantum channel does not introduce further errors, the syndrome measured at the decoder will be $\hat{\Syndrome}_i=\Syndrome(\M{P}_i)=\Syndrome_i$. Therefore, 
a sequence of syndromes, carrying classical bits of information, is sent through this error free \ac{PSC}. Moreover, the $\M{P}_i$ can be determined from the measured syndrome and the state $\ket{\psi_i}$  can be restored by applying $\M{P}_i^\dag$ to the received state $\M{P}_i\ket{\psi_i}$. 
 The proposed piggybacking method enables sending $n-k$ classical bits per q-codeword, without additional qubits and without destroying quantum superposition. 

A possible application consists in adding, to a group of q-codewords, control data in terms of classical bits that one can read and write without altering the quantum information. Fig.~\ref{fig:QPexamples}(a) illustrates an application example, in which a unique pattern of intentional errors is sent over the \ac{PSC} for \acl{FS}, similar to the classical case   \cite{SuwChiWin:J08}. 
Another application is annotation of a quantum stream, 
as reported in Fig.~\ref{fig:QPexamples}(b) and Fig.~\ref{fig:QPexamples}(c). 
\begin{figure}[h]
\vspace{0.2cm}
\begin{center}{\scriptsize
\begin{tabular}{*{8}{p{1.05cm}}}
	$\ket{\psi_1}$ & $\ket{\psi_2}$ & $\ket{\psi_3}$ & $\ket{\psi_4}$ & $\M{Q}^{(4)}\ket{\psi_5}$ & $\M{Q}^{(2)}\ket{\psi_6}$ & $\M{Q}^{(3)}\ket{\psi_7}$\\
	$\Syndrome^{(1)}$& $\Syndrome^{(1)}$& $\Syndrome^{(1)}$& $\Syndrome^{(1)}$ & $\Syndrome^{(4)}$ & $\Syndrome^{(2)}$ & $\Syndrome^{(3)}$ \\
\end{tabular}
\\ \smallskip $(a)$ \\ \bigskip
\begin{tabular}{c || c c c c c c}
$\ket{\psi_i}$ & $\cdots$ & $\ket{\psi_1}$& $\ket{\psi_2}$& $\ket{\psi_3}$& $\ket{\psi_4}$& $\cdots$ \\
$\Syndrome_i$ & $\cdots$ & $\Syndrome^{(3)}$& $\Syndrome^{(1)}$& $\Syndrome^{(11)}$& $\Syndrome^{(7)}$& $\cdots$ \\
$\M{P}_i \ket{\psi_i}$ & $\cdots$ & $\M{Q}^{(3)}\ket{\psi_1}$& $\M{Q}^{(1)}\ket{\psi_2}$& $\M{Q}^{(11)}\ket{\psi_3}$& $\M{Q}^{(7)}\ket{\psi_4}$& $\cdots$ \\
$\hat{\Syndrome}_i$ & $\cdots$ & $\Syndrome^{(3)}$& $\Syndrome^{(1)}$& $\Syndrome^{(11)}$& $\Syndrome^{(7)}$& $\cdots$ \\
\end{tabular}
\\ \smallskip $(b)$ \\ \bigskip
\begin{tabular}{ c c c c c}
	 $\M{Q}^{(2)}\ket{\psi_1}$& $\M{Q}^{(4)}\ket{\psi_2}$ & $\M{Q}^{(1)}\ket{\psi_3}$ & $\M{Q}^{(3)}\ket{\psi_4}$ & $\M{Q}^{(1)}\ket{\psi_5}$ \\ 
	 $-1 +1 $& $+1 -1 $& $+1 +1 $& $-1 -1 $& $+1 +1$ \\
\end{tabular}
\\ \smallskip $(c)$ \\ 
}\end{center}
\vspace{-0.2cm}
\caption{Examples of piggybacking on a noiseless channel.  
$(a)$ Piggybacking synchronization patterns on a quantum stream, frames composed of $7$ q-codewords. The synchronization word pattern used in this example is $\Syndrome^{(4)},\Syndrome^{(2)},\Syndrome^{(3)}$, and thus synchronization is obtained with three intentional errors on the last three q-codewords.
 $(b)$  
 Piggybacking the classical information $\Syndrome^{(3)}, \Syndrome^{(1)}, \ldots $ on the quantum stream $\ket{\psi_1},\ket{\psi_2}, \ldots$. 
 $(c)$ 
 Piggybacking $10$ classical bits of information over a quantum packet composed of $5$ q-codewords, assuming each q-codeword $\ket{\psi_i}=\alpha_i \ket{000} + \beta_i \ket{111}$ is from a repetition $[[3,1]]$ \ac{QECC}. 
	 Here 
 $\M{Q}^{(1)}=$ no error, $\M{Q}^{(2)}=$ bit-flip on the first qubit, $\M{Q}^{(3)}=$ bit-flip on the second qubit, $\M{Q}^{(4)}=$ bit-flip on the third qubit. }
\label{fig:QPexamples}
\end{figure}

\subsection{Piggybacking over a noisy quantum channel}
\begin{figure*}[t]
	\begin{center}
		\includegraphics[width=1.6\columnwidth]{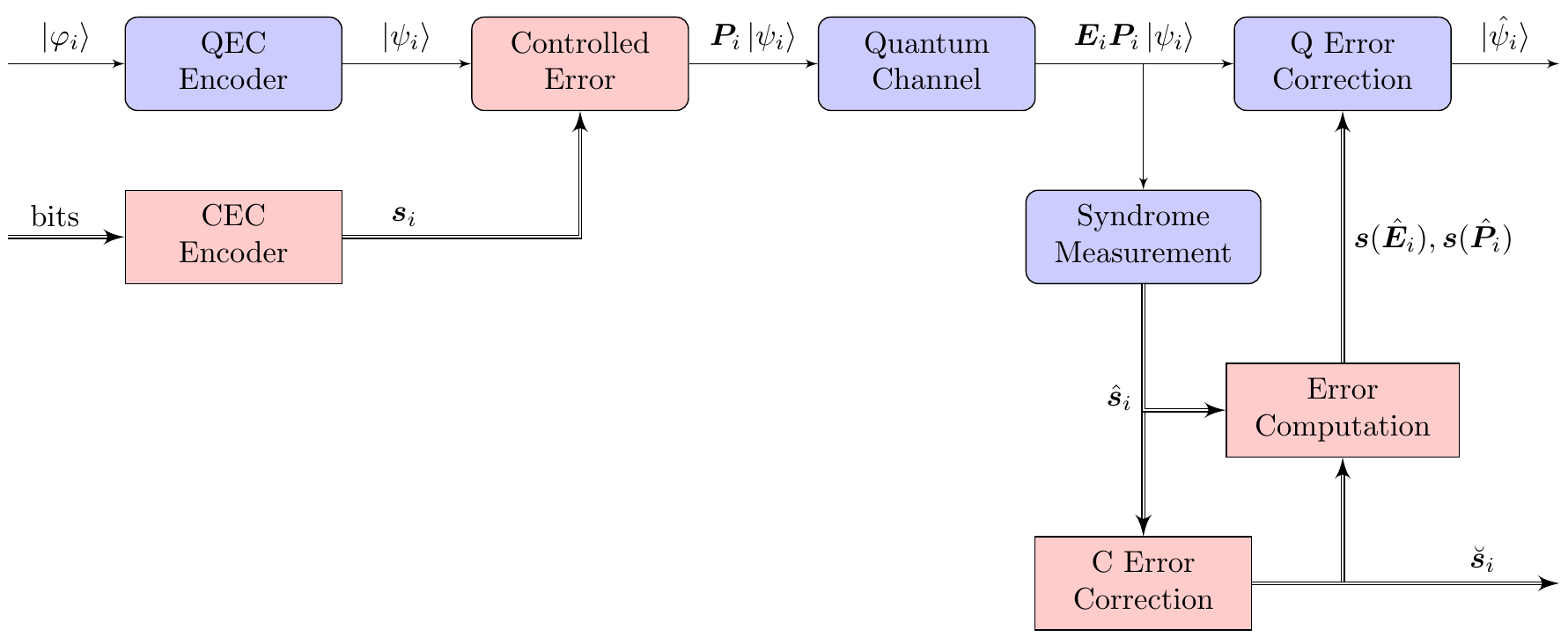}
	\end{center}
	\caption{Quantum piggybacking by intentional errors, noisy quantum channel. The blocks added to realize the piggyback classical channel are in red.}
	\label{fig:piggyblockdiagram}
\end{figure*}

We now consider a noisy quantum channel and, as in the previous case, intentionally apply $\M{P}_i\in \mathcal{Q}$ on the $i-$th q-codeword. 
If the quantum channel introduces an error $\M{E}_i \in \mathcal{G}_n$, then the measured syndrome will be $\hat{\Syndrome}_i=\Syndrome(\M{E}_i \M{P}_i)$ corresponding to the composite operator $\M{E}_i \M{P}_i$  (see Fig.~\ref{fig:piggyblockdiagram}).  
Therefore, the \ac{PSC} can be seen as a classical channel with errors. 
To cope with the effects of quantum channel errors on the  \ac{PSC}, we envision the use of  \acp{CECC}, with alphabet $\mathcal{S}$ for the encoded symbols.  
In particular, classical information is encoded at the transmitter so that a vector of transmitted syndromes $(\Syndrome_1, \Syndrome_2, \ldots, \Syndrome_{N})$ is a length $N$ c-codeword. The   \ac{CECC} can be any one of the classical codes, including BCH, RS, Convolutional, LDPC, and Turbo codes \cite{RL09}.  
A classical error correction block at the receiver aims to correct the errors due to the quantum channel; at its output the syndromes $\breve{\Syndrome}_i$ are equal to the transmitted syndromes $\Syndrome_i$ in case of successful correction (see Fig.~\ref{fig:piggyblockdiagram}). 
Indeed, by employing \acp{CECC} with rates below the \ac{PSC} capacity, errors on the received syndromes caused by the quantum channel can be corrected with probability arbitrarily close to one \cite{CovTho:06}. 
Therefore, applications similar to those for the noiseless case in Fig.~\ref{fig:QPexamples} are also possible for the noisy case.

The presence of both intentional  and unintentional errors, described respectively by the operators $\M{P}_i$ and $\M{E}_i$, has to be taken into account in the quantum error correction process. Recall that the $\M{P}_i$ is correctable by design; nevertheless, the combined error may not be in the set of correctable errors, i.e., $\M{E}_i \M{P}_i \notin \mathcal{Q}$ even if $\M{E}_i \in \mathcal{Q}$.  
For example, one intentional error  together with  one quantum channel error on different qubits of the same q-codeword would produce an uncorrectable combined error for a   \ac{QECC}  with single qubit error correction capability.  
The \ac{CECC} can also help alleviating this problem, in addition to protecting the \ac{PSC} from quantum channel errors as discussed before. 
In fact, if classical error correction on the \ac{PSC} is successful, the intentional error with operator $\M{P}_i$ is known and the quantum channel error with operator $\M{E}_i \in \mathcal{Q}$  can be determined by observing that  the measured syndrome is $\hat{\Syndrome}_i=\Syndrome(\M{E}_i \M{P}_i)=\Syndrome(\M{E}_i) \circ \Syndrome(\M{P}_i)$, where $\circ$ denotes the Hadamard product. 
In fact, $\Syndrome(\M{E}_i \M{P}_i)=+1$ if $\M{E}_i \M{P}_i$ commutes with $\M{G}_i$, which happens when $\M{E}_i$ and $\M{P}_i$ both commute or both anticommute with $\M{G}_i$. 
Since the syndrome elements are $\pm 1$, it follows {also} that $\Syndrome(\M{E}_i)=\hat{\Syndrome}_i \circ \Syndrome(\M{P}_i)$. 

The classical decoder provides $\breve{\Syndrome}_i= \Syndrome(\hat{\M{P}}_i)$  where $\hat{\M{P}}_i$ is the operator corresponding to the  estimated intentional error.  
Therefore, the error computation block in Fig.~\ref{fig:piggyblockdiagram} infers $E_{i}$ by computing $ \hat{\Syndrome}_i \circ \breve{\Syndrome}_i=  \Syndrome(\hat{\M{E}}_i)$, where $\hat{\M{E}}_i$  is the operator corresponding to the estimated quantum channel error. Then, the quantum error correction block recovers the quantum state from the  composite error by applying $\hat{\M{P}}_i^\dag\hat{\M{E}}_i^\dag$. 
Notice that, with the proposed method, the capability of the \ac{QECC} is not affected by piggybacking, as long as the errors in the \ac{PSC} are successfully corrected by the \ac{CECC}. 
%

\section{Capacity of the Piggyback Syndrome Channel}%
As observed, if the quantum channel is noiseless, $n-k$ $\text{[bits/q-codeword]}$ can be sent over the \ac{PSC}. 
On the other hand, if the quantum channel is noisy, the capacity of the \ac{PSC} depends on the statistics of the quantum channel errors. 
In particular, if the quantum error process is memoryless (i.e., errors $\M{E}_i$ and $\M{E}_j$ are independent for $i \neq j$), the \ac{PSC} is a classical \acl{DMC}; its  mutual information is determined by the transition probability together with the probability distribution $p(\Syndrome)$ of the input $\SyndromeRV$.  
The capacity in $\text{[bits/q-codeword]}$ is 
\begin{align}\label{eq:cesc}
C_\mathrm{PSC}= 
\max_{p(\Syndrome)} \big\{ H(\SyndromeRV)-H(\SyndromeRV|\hat{\SyndromeRV})\big\}  
\end{align}
where $H(\SyndromeRV)$ is the Shannon entropy \cite{CovTho:06}. 

Define the probability that the measured syndrome at the receiver is different from the transmitted syndrome as
\begin{equation}\label{eq:defppsc}
p_\mathrm{PSC}=\Pr\big\{\hat{\SyndromeRV}\neq \SyndromeRV\big\} \,.
\end{equation}
Consider a quantum channel error that maps the transmitted syndrome into one of the remaining $2^{n-k}-1$ syndromes with equal probability, which is the worst case for the capacity. 
Then, the \ac{PSC} is an $m$-ary symmetric channel ($m=2^{n-k}$) with transition probabilities 
\begin{align*}
\Pr\big\{\hat{\SyndromeRV} = \Syndrome^{(j)} | \SyndromeRV = \Syndrome^{(i)}\big\}  = 
 \begin{cases} 1-p_\mathrm{PSC} & j=i \\
 \frac{p_\mathrm{PSC}}{2^{n-k}-1} & j \neq i \,.
 \end{cases} 
\end{align*}
From \eqref{eq:cesc}, the capacity results in
\begin{align}\label{cpsc}
C_\mathrm{PSC}= n-k-{h}(p_\mathrm{PSC})-p_\mathrm{PSC} \log_2 \left(2^{n-k}-1\right) 
\end{align}
where ${h}(p)=-p \log_2 p - (1-p) \log_2 (1-p)$ is the binary entropy function. 

\begin{figure}[t]
	\vspace{-3pt}
	\begin{center}
		\includegraphics[width=1.18\columnwidth]{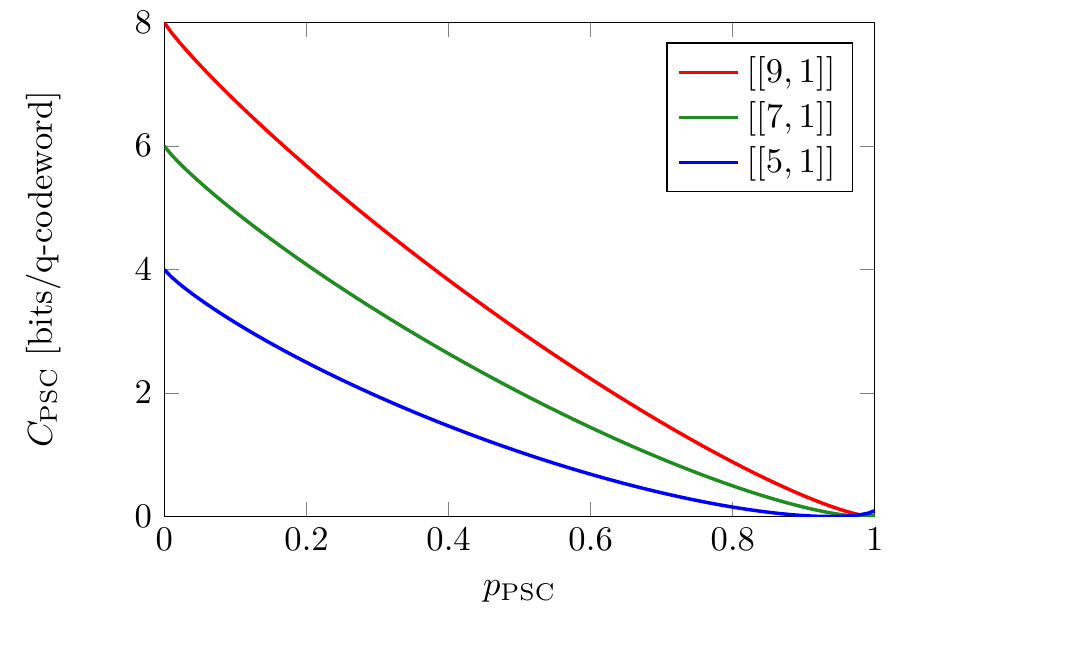}
	\end{center}
	\vspace{-20pt}
	\caption{Capacity of the \acl{PSC} as a function of $p_\mathrm{PSC}$ for different \acp{QECC}.}
	\label{fig:CPSC}
\end{figure}

\begin{figure}[t]
	\begin{center}
		\includegraphics[width=1.16\columnwidth]{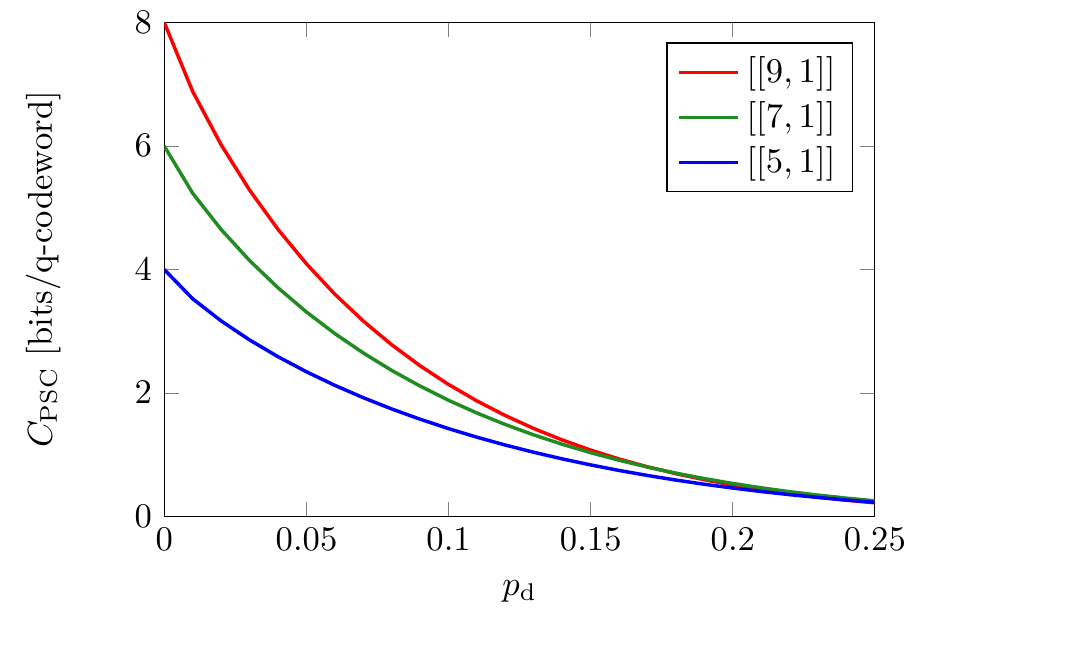}
	\end{center}
	\vspace{-22pt}
	\caption{Lower bound on the capacity of the \acl{PSC} for a quantum depolarizing channel as a function of $p_\mathrm{d}$ for different \acp{QECC}.
	}
	\label{fig:CPD}
\end{figure}

The capacity in \eqref{cpsc} is completely characterized by $p_\mathrm{PSC}$, which depends on the characteristics of the quantum channel.  
An upper bound for $p_\mathrm{PSC}$ is the probability that the quantum channel introduces an error on a q-codeword,  $\Pr\big\{\M{E}_i \neq \M{I} \big\}$, 
since this accounts also for the undetectable quantum errors for which 
 $\Syndrome(\M{E}_i \M{P}_i)=\Syndrome(\M{P}_i)$. %
For example, consider a memoryless quantum depolarizing channel, where each qubit is subject to no error (operator $\M{I}$) with probability $1-p_\mathrm{d}$, or to errors of type $\M{X}, \M{Y}$ and $\M{Z}$ each with probability $p_\mathrm{d}/3$ \cite{NieChu:B10}. 
For a q-codeword of $n$ qubits, 
$p_\mathrm{PSC} < \Pr\big\{\M{E}_i \neq \M{I} \big\}=1-\left(1-p_\mathrm{d}\right)^n$, which together with \eqref{cpsc} provides a lower bound on the capacity.  
Fig.~\ref{fig:CPSC} shows $C_\mathrm{PSC}$ as a function of $p_\mathrm{PSC}$ for the $[[5,1]], [[7,1]]$, and $[[9,1]]$ \acp{QECC} with single qubit error correction capability \cite{NieChu:B10}. 
Notice that for $p_\mathrm{PSC}=0$ the capacity is that of the noiseless case, i.e., $n-k$ $\text{[bits/q-codeword]}$. For noisy channels with $p_\mathrm{PSC}=0.1$ the loss in capacity is of around one bit. 
Fig.~\ref{fig:CPD} shows the lower bound on the capacity of a memoryless quantum depolarizing channel as a function of $p_\mathrm{d}$ for the same \acp{QECC}. 
For a given value of $p_\mathrm{d}$, characteristic of the quantum depolarizing channel, it is possible to determine what is the minimum guaranteed capacity of the \ac{PSC} for different quantum codes. 
\section{Impact of piggybacking on the \ac{QECC} capability}As observed, the \ac{PSC} does not affect the capability of the \ac{QECC} if the errors on the measured syndromes are successfully corrected by the classical error correction block in Fig.~\ref{fig:piggyblockdiagram}. If the classical error correction fails, then $\breve{\Syndrome}_i\neq \Syndrome_i$ for some $i$, causing also a failure in the quantum error correction block. 
Thus, the probability of a q-codeword error due to piggybacking is equal to  the probability of residual syndrome error after decoding 
\begin{equation}\label{eq:pqep}
p_\mathrm{QEP}=\Pr\big\{\breve{\Syndrome}_i\neq \Syndrome_i\big\}
\end{equation} 
which depends on both the quantum channel and the specific \ac{CECC} used. 
\begin{figure}[t]
	\begin{center}\vspace{-0.03cm}
		\includegraphics[width=1.16\columnwidth]{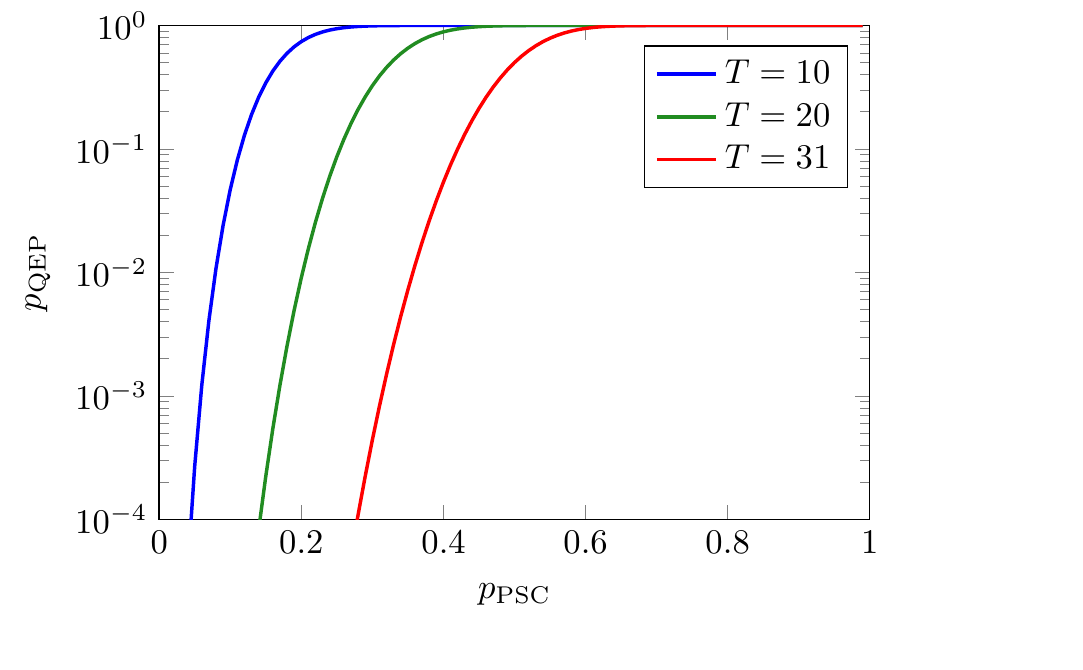}
	\end{center}
	\vspace{-24pt}
	\caption{Upper bound on the probability of a q-codeword error due to piggybacking, $p_\mathrm{QEP}$, as a function of $p_\mathrm{PSC}$ for different correction capabilities $T$.}
	\label{fig:peresid71}
\end{figure} 

The case of 
 a quantum link employing an $[[n,k]]$ quantum code and a nonbinary $(N,K)$ classical code over the Galois field $\GF{2^{n-k}}$ is illustrated in the following. With this choice,  
  each c-codeword symbol is mapped into a syndrome. The encoder then takes $K$ information syndromes (i.e., $(n-k) K$ classical bits) and produces a c-codeword of $N$ syndromes, resulting in $(n-k) K/N$ [bits/q-codeword].    
Thus, the allowed code rates can be determined using \eqref{cpsc} as
\begin{align}\label{eq:caprateRS}
\frac{K}{N} <  1-\frac{{h}(p_\mathrm{PSC})+p_\mathrm{PSC} \log_2 \left(2^{n-k}-1\right)}{n-k}\,.
\end{align}

To characterize the $p_\mathrm{QEP}$ in \eqref{eq:pqep}, 
consider a Reed Solomon code RS$(N,K)$ over $\GF{2^{n-k}}$ with length $N=2^{n-k}-1$. Since it is a maximum distance separable code, 
  the RS$(N, N-2T)$ can correct up to $T$ errors per c-codeword \cite{RL09}. For this code, the probability of a q-codeword error due to piggybacking is upper bounded by
\begin{align}\label{eq:pepscresidual}
p_\mathrm{QEP} \leq \hspace{-0.1cm}\sum_{\ell=T+1}^{N} {N \choose \ell} p_\mathrm{PSC}^\ell \left(1-p_\mathrm{PSC}\right)^{N-\ell} \,.
\end{align}

Fig.~\ref{fig:peresid71} shows the upper bound as a function of $p_\mathrm{PSC}$ when using the $[[7,1]]$ \ac{QECC} together with RS$(63,63-2T)$ codes over $\GF{2^6}$ for the \ac{PSC}.  
It can be seen that with the RS$(63,23)$ code, which can correct up to $T=20$ erroneous syndromes per c-codeword, the \ac{PSC} has a negligible impact on the quantum stream ($p_\mathrm{QEP} < 10^{-4}$) whenever 
 $p_\mathrm{PSC}<0.15$. Notice that the quantum decoder will experience a delay of $N = 63$ q-codewords. 

The impact of piggybacking on quantum streams can be reduced for a given quantum channel (equivalently, noisier quantum channels can be considered for a given maximum tolerable q-codeword error probability) 
 by using more powerful \acp{CECC}, without consuming additional quantum resources. 
However, it is important to observe that using longer c-codewords results in larger delay for the quantum error correction. 
Therefore, a performance--delay trade-off has to be accounted for in designing  
\acp{CECC} to control the impact of piggybacking on quantum streams.

\section{Conclusion}
We put forth a method to piggyback up to $n-k$ classical bits on top of each q-codeword of an $[[n,k]]$ quantum code. 
Such piggyback operation exploits the syndromes of a quantum code by leveraging classical codes 
according to a performance--delay trade-off. The proposed method enables new capabilities, even for noisy quantum channels, without consuming additional quantum resources and without disturbing the quantum stream.

\bibliographystyle{apsrev} 
\bibliography{StringDefinitions,IEEEabrv,quantumCU,BiblioFS,BiblioCV,BiblioMCCV}

\end{document}

%% file: acronymsMC.tex
\begin{acronym}
\scriptsize
\acro{AcR}{autocorrelation receiver}
\acro{ACF}{autocorrelation function}
\acro{ADC}{analog-to-digital converter}
\acro{AWGN}{additive white Gaussian noise}
\acro{BCH}{Bose Chaudhuri Hocquenghem}
\acro{BEP}{bit error probability}
\acro{BFC}{block fading channel}
\acro{BPAM}{binary pulse amplitude modulation}
\acro{BPPM}{binary pulse position modulation}
\acro{BPSK}{binary phase shift keying}
\acro{BPZF}{bandpass zonal filter}
\acro{CD}{cooperative diversity}
\acro{CDF}{cumulative distribution function}
\acro{CCDF}{complementary cumulative distribution function}
\acro{CDMA}{code division multiple access}
\acro{c.d.f.}{cumulative distribution function}
\acro{c.c.d.f.}{complementary cumulative distribution function}
\acro{ch.f.}{characteristic function}
\acro{CIR}{channel impulse response}
\acro{CR}{cognitive radio}
\acro{CSI}{channel state information}
\acro{DAA}{detect and avoid}
\acro{DAB}{digital audio broadcasting}
\acro{DS}{direct sequence}
\acro{DS-SS}{direct-sequence spread-spectrum}
\acro{DTR}{differential transmitted-reference}
\acro{DVB-T}{digital video broadcasting\,--\,terrestrial}
\acro{DVB-H}{digital video broadcasting\,--\,handheld}
\acro{ECC}{European Community Commission}
\acro{ELP}{equivalent low-pass}
\acro{FCC}{Federal Communications Commission}
\acro{FEC}{forward error correction}
\acro{FFT}{fast Fourier transform}
\acro{FH}{frequency-hopping}
\acro{FH-SS}{frequency-hopping spread-spectrum}
\acro{GA}{Gaussian approximation}
\acro{GPS}{Global Positioning System}
\acro{HAP}{high altitude platform}
\acro{i.i.d.}{independent, identically distributed}
\acro{IFFT}{inverse fast Fourier transform}
\acro{IR}{impulse radio}
\acro{ISI}{intersymbol interference}
\acro{LEO}{low earth orbit}
\acro{LOS}{line-of-sight}
\acro{BSC}{binary symmetric channel}
\acro{MB}{multiband}
\acro{MC}{multicarrier}
\acro{MF}{matched filter}
\acro{m.g.f.}{moment generating function}
\acro{MI}{mutual information}
\acro{MIMO}{multiple-input multiple-output}
\acro{MISO}{multiple-input single-output}
\acro{MRC}{maximal ratio combiner}
\acro{MMSE}{minimum mean-square error}
\acro{M-QAM}[$M$-QAM]{$M$-ary quadrature amplitude modulation}
\acro{M-PSK}[${M}$-PSK]{$M$-ary phase shift keying}
\acro{MUI}{multi-user interference}
\acro{NB}{narrowband}
\acro{NBI}{narrowband interference}
\acro{NLOS}{non-line-of-sight}
\acro{NTIA}{National Telecommunications and Information Administration}
\acro{OC}{optimum combining}
\acro{OFDM}{orthogonal frequency-division multiplexing}
\acro{p.d.f.}{probability distribution function}
\acro{PAM}{pulse amplitude modulation}
\acro{PAR}{peak-to-average ratio}
\acro{PDP}{power dispersion profile}
\acro{p.m.f.}{probability mass function}
\acro{PN}{pseudo-noise}
\acro{PPM}{pulse position modulation}
\acro{PRake}{Partial Rake}
\acro{PSD}{power spectral density}
\acro{PSK}{phase shift keying}
\acro{QAM}{quadrature amplitude modulation}
\acro{QKD}{quantum key distribution}
\acro{QPSK}{quadrature phase shift keying}
\acro{8-PSK}[$8$-PSK]{$8$-phase shift keying}
\acro{r.v.}{random variable}
\acro{R.V.}{random vector}
\acro{SEP}{symbol error probability}
\acro{SIMO}{single-input multiple-output}
\acro{SIR}{signal-to-interference ratio}
\acro{SISO}{single-input single-output}
\acro{SINR}{signal-to-interference plus noise ratio}
\acro{SNR}{signal-to-noise ratio}
\acro{SS}{spread spectrum}
\acro{TH}{time-hopping}
\acro{ToA}{time-of-arrival}
\acro{TR}{transmitted-reference}
\acro{UAV}{unmanned aerial vehicle}
\acro{UWB}{ultrawide band}
\acro{UWBcap}[UWB]{Ultrawide band}
\acro{WLAN}{wireless local area network}
\acro{WMAN}{wireless metropolitan area network}
\acro{WPAN}{wireless personal area network}
\acro{WSN}{wireless sensor network}
\acro{WSS}{wide-sense stationary}

\acro{SW}{sync word}
\acro{FS}{frame synchronization}
\acro{FSsmall}[FS]{frame synchronization}
\acro{BSC}{binary symmetric channels}
\acro{LRT}{likelihood ratio test}
\acro{GLRT}{generalized likelihood ratio test}
\acro{LLRT}{log-likelihood ratio test}
\acro{LLR}{log-likelihood ratio}
\acro{$P_{EM}$}{probability of emulation, or false alarm}
\acro{$P_{MD}$}{probability of missed detection}
\acro{ROC}{receiver operating characteristic}
\acro{AUB}{asymptotic union bound}
\acro{RDL}{"random data limit"}
\acro{PSEP}{pairwise synchronization error probability}

\acro{SCM}{sample covariance matrix}

\acro{QECC}{quantum error correcting code}
\acro{CECC}{classical error correcting code}
\acro{DMC}{discrete memoryless channel}
\acro{PSC}{piggyback syndrome channel}
\acro{QN}{quantum network}
\acro{IP}{Internet Protocol}

\end{acronym}

%% file: QuantumPiggyR1.bbl
\begin{thebibliography}{24}
\expandafter\ifx\csname natexlab\endcsname\relax\def\natexlab#1{#1}\fi
\expandafter\ifx\csname bibnamefont\endcsname\relax
  \def\bibnamefont#1{#1}\fi
\expandafter\ifx\csname bibfnamefont\endcsname\relax
  \def\bibfnamefont#1{#1}\fi
\expandafter\ifx\csname citenamefont\endcsname\relax
  \def\citenamefont#1{#1}\fi
\expandafter\ifx\csname url\endcsname\relax
  \def\url#1{\texttt{#1}}\fi
\expandafter\ifx\csname urlprefix\endcsname\relax\def\urlprefix{URL }\fi
\providecommand{\bibinfo}[2]{#2}
\providecommand{\eprint}[2][]{\url{#2}}

\bibitem[{\citenamefont{{P. Zoller et al.}}(2005)}]{Zol:05}
\bibinfo{author}{\bibnamefont{{P. Zoller et al.}}}, \bibinfo{journal}{Eur.
  Phys. J. D} \textbf{\bibinfo{volume}{36}}, \bibinfo{pages}{203}
  (\bibinfo{year}{2005}).

\bibitem[{\citenamefont{Kimble}(2008)}]{kim:08}
\bibinfo{author}{\bibfnamefont{H.~J.} \bibnamefont{Kimble}},
  \bibinfo{journal}{Nature} \textbf{\bibinfo{volume}{453}},
  \bibinfo{pages}{1023} (\bibinfo{year}{2008}).

\bibitem[{\citenamefont{Van~Meter and Touch}(2013)}]{Van:13}
\bibinfo{author}{\bibfnamefont{R.}~\bibnamefont{Van~Meter}} \bibnamefont{and}
  \bibinfo{author}{\bibfnamefont{J.}~\bibnamefont{Touch}},
  \bibinfo{journal}{{IEEE} Commun. Mag.} \textbf{\bibinfo{volume}{51}},
  \bibinfo{pages}{64} (\bibinfo{year}{2013}).

\bibitem[{\citenamefont{Dai et~al.}(2020)\citenamefont{Dai, Peng, and
  Win}}]{DaiPenWin:J20b}
\bibinfo{author}{\bibfnamefont{W.}~\bibnamefont{Dai}},
  \bibinfo{author}{\bibfnamefont{T.}~\bibnamefont{Peng}}, \bibnamefont{and}
  \bibinfo{author}{\bibfnamefont{M.~Z.} \bibnamefont{Win}},
  \bibinfo{journal}{{IEEE} J. Sel. Areas Commun.}
  \textbf{\bibinfo{volume}{38}}, \bibinfo{pages}{540} (\bibinfo{year}{2020}).

\bibitem[{\citenamefont{Wehner et~al.}(2018)\citenamefont{Wehner, Elkouss, and
  Hanson}}]{Weh:18}
\bibinfo{author}{\bibfnamefont{S.}~\bibnamefont{Wehner}},
  \bibinfo{author}{\bibfnamefont{D.}~\bibnamefont{Elkouss}}, \bibnamefont{and}
  \bibinfo{author}{\bibfnamefont{R.}~\bibnamefont{Hanson}},
  \bibinfo{journal}{Science} \textbf{\bibinfo{volume}{362}}
  (\bibinfo{year}{2018}).

\bibitem[{\citenamefont{Grumbling and Horowitz}(2019)}]{NAP:19}
\bibinfo{editor}{\bibfnamefont{E.}~\bibnamefont{Grumbling}} \bibnamefont{and}
  \bibinfo{editor}{\bibfnamefont{M.}~\bibnamefont{Horowitz}}, eds.,
  \emph{\bibinfo{title}{Quantum Computing: Progress and Prospects}}
  (\bibinfo{publisher}{The National Academies Press},
  \bibinfo{address}{Washington, DC}, \bibinfo{year}{2019}).

\bibitem[{OSW(2018)}]{OSW18}
\emph{\bibinfo{title}{Quantum Networks for Open Science Workshop}}
  (\bibinfo{publisher}{Office of Science US Department of Energy},
  \bibinfo{address}{Rockville, MD, USA}, \bibinfo{year}{2018}).

\bibitem[{\citenamefont{Suwansantisuk et~al.}(2008)\citenamefont{Suwansantisuk,
  Chiani, and Win}}]{SuwChiWin:J08}
\bibinfo{author}{\bibfnamefont{W.}~\bibnamefont{Suwansantisuk}},
  \bibinfo{author}{\bibfnamefont{M.}~\bibnamefont{Chiani}}, \bibnamefont{and}
  \bibinfo{author}{\bibfnamefont{M.~Z.} \bibnamefont{Win}},
  \bibinfo{journal}{{IEEE} J. Sel. Areas Commun.}
  \textbf{\bibinfo{volume}{26}}, \bibinfo{pages}{52} (\bibinfo{year}{2008}).

\bibitem[{\citenamefont{Nielsen and Chuang}(2010)}]{NieChu:B10}
\bibinfo{author}{\bibfnamefont{M.~A.} \bibnamefont{Nielsen}} \bibnamefont{and}
  \bibinfo{author}{\bibfnamefont{I.~L.} \bibnamefont{Chuang}},
  \emph{\bibinfo{title}{Quantum Computation and Quantum Information}}
  (\bibinfo{publisher}{Cambridge University Press},
  \bibinfo{address}{Cambridge, UK}, \bibinfo{year}{2010}),
  \bibinfo{edition}{2nd} ed.

\bibitem[{\citenamefont{Qi}(2016)}]{Qi:16}
\bibinfo{author}{\bibfnamefont{B.}~\bibnamefont{Qi}}, \bibinfo{journal}{Phys.
  Rev. A} \textbf{\bibinfo{volume}{94}}, \bibinfo{pages}{042340}
  (\bibinfo{year}{2016}).

\bibitem[{\citenamefont{Wu et~al.}(2019)\citenamefont{Wu, Wang, Liao, Zhong,
  and Guo}}]{WuWanLiaZhoGuo:19}
\bibinfo{author}{\bibfnamefont{X.}~\bibnamefont{Wu}},
  \bibinfo{author}{\bibfnamefont{Y.}~\bibnamefont{Wang}},
  \bibinfo{author}{\bibfnamefont{Q.}~\bibnamefont{Liao}},
  \bibinfo{author}{\bibfnamefont{H.}~\bibnamefont{Zhong}}, \bibnamefont{and}
  \bibinfo{author}{\bibfnamefont{Y.}~\bibnamefont{Guo}},
  \bibinfo{journal}{Entropy} \textbf{\bibinfo{volume}{21}},
  \bibinfo{pages}{333} (\bibinfo{year}{2019}).

\bibitem[{\citenamefont{Kumar et~al.}(2019)\citenamefont{Kumar, Wonfor, Penty,
  Spiller, and White}}]{KumWonPenSpiWhi:19}
\bibinfo{author}{\bibfnamefont{R.}~\bibnamefont{Kumar}},
  \bibinfo{author}{\bibfnamefont{A.}~\bibnamefont{Wonfor}},
  \bibinfo{author}{\bibfnamefont{R.}~\bibnamefont{Penty}},
  \bibinfo{author}{\bibfnamefont{T.}~\bibnamefont{Spiller}}, \bibnamefont{and}
  \bibinfo{author}{\bibfnamefont{I.}~\bibnamefont{White}},
  \bibinfo{journal}{Scientific reports} \textbf{\bibinfo{volume}{9}},
  \bibinfo{pages}{1} (\bibinfo{year}{2019}).

\bibitem[{\citenamefont{Fujiwara}(2013)}]{Fuj:13}
\bibinfo{author}{\bibfnamefont{Y.}~\bibnamefont{Fujiwara}},
  \bibinfo{journal}{{IEEE} Trans. Inf. Theory} \textbf{\bibinfo{volume}{59}},
  \bibinfo{pages}{6796} (\bibinfo{year}{2013}).

\bibitem[{\citenamefont{Shor}(1995)}]{Sho:95}
\bibinfo{author}{\bibfnamefont{P.~W.} \bibnamefont{Shor}},
  \bibinfo{journal}{Phys. Rev. A} \textbf{\bibinfo{volume}{52}},
  \bibinfo{pages}{2493} (\bibinfo{year}{1995}).

\bibitem[{\citenamefont{Calderbank and Shor}(1996)}]{CalSho:96}
\bibinfo{author}{\bibfnamefont{A.~R.} \bibnamefont{Calderbank}}
  \bibnamefont{and} \bibinfo{author}{\bibfnamefont{P.~W.} \bibnamefont{Shor}},
  \bibinfo{journal}{Phys. Rev. A} \textbf{\bibinfo{volume}{54}},
  \bibinfo{pages}{1098} (\bibinfo{year}{1996}).

\bibitem[{\citenamefont{Steane}(1996)}]{Ste:96}
\bibinfo{author}{\bibfnamefont{A.~M.} \bibnamefont{Steane}},
  \bibinfo{journal}{Phys. Rev. Lett.} \textbf{\bibinfo{volume}{77}},
  \bibinfo{pages}{793} (\bibinfo{year}{1996}).

\bibitem[{\citenamefont{Laflamme et~al.}(1996)\citenamefont{Laflamme, Miquel,
  Paz, and Zurek}}]{LafMiqPazZur:96}
\bibinfo{author}{\bibfnamefont{R.}~\bibnamefont{Laflamme}},
  \bibinfo{author}{\bibfnamefont{C.}~\bibnamefont{Miquel}},
  \bibinfo{author}{\bibfnamefont{J.-P.} \bibnamefont{Paz}}, \bibnamefont{and}
  \bibinfo{author}{\bibfnamefont{W.~H.} \bibnamefont{Zurek}},
  \bibinfo{journal}{Phys. Rev. Lett.} \textbf{\bibinfo{volume}{77:198}}
  (\bibinfo{year}{1996}).

\bibitem[{\citenamefont{Gottesman}(1996)}]{Got:96}
\bibinfo{author}{\bibfnamefont{D.}~\bibnamefont{Gottesman}},
  \bibinfo{journal}{Phys. Rev. A} \textbf{\bibinfo{volume}{54:1862}}
  (\bibinfo{year}{1996}).

\bibitem[{\citenamefont{Ekert and Macchiavello}(1996)}]{EkeMac:96}
\bibinfo{author}{\bibfnamefont{A.}~\bibnamefont{Ekert}} \bibnamefont{and}
  \bibinfo{author}{\bibfnamefont{C.}~\bibnamefont{Macchiavello}},
  \bibinfo{journal}{Phys. Rev. Lett.} \textbf{\bibinfo{volume}{77}},
  \bibinfo{pages}{2585} (\bibinfo{year}{1996}).

\bibitem[{\citenamefont{Fletcher et~al.}(2007)\citenamefont{Fletcher, Shor, and
  Win}}]{FleShoWin:J07}
\bibinfo{author}{\bibfnamefont{A.~S.} \bibnamefont{Fletcher}},
  \bibinfo{author}{\bibfnamefont{P.~W.} \bibnamefont{Shor}}, \bibnamefont{and}
  \bibinfo{author}{\bibfnamefont{M.~Z.} \bibnamefont{Win}},
  \bibinfo{journal}{Phys. Rev. A} \textbf{\bibinfo{volume}{75}},
  \bibinfo{pages}{012338} (\bibinfo{year}{2007}).

\bibitem[{\citenamefont{Gottesman}(2009)}]{gottesman2009introduction}
\bibinfo{author}{\bibfnamefont{D.}~\bibnamefont{Gottesman}}, in
  \emph{\bibinfo{booktitle}{Proceedings of Symposia in Applied Mathematics}}
  (\bibinfo{year}{2009}), vol.~\bibinfo{volume}{68}, pp.
  \bibinfo{pages}{13--58}.

\bibitem[{\citenamefont{{Babar} et~al.}(2019)\citenamefont{{Babar}, {Chandra},
  {Nguyen}, {Botsinis}, {Alanis}, {Ng}, and {Hanzo}}}]{BabHan:19}
\bibinfo{author}{\bibfnamefont{Z.}~\bibnamefont{{Babar}}},
  \bibinfo{author}{\bibfnamefont{D.}~\bibnamefont{{Chandra}}},
  \bibinfo{author}{\bibfnamefont{H.~V.} \bibnamefont{{Nguyen}}},
  \bibinfo{author}{\bibfnamefont{P.}~\bibnamefont{{Botsinis}}},
  \bibinfo{author}{\bibfnamefont{D.}~\bibnamefont{{Alanis}}},
  \bibinfo{author}{\bibfnamefont{S.~X.} \bibnamefont{{Ng}}}, \bibnamefont{and}
  \bibinfo{author}{\bibfnamefont{L.}~\bibnamefont{{Hanzo}}},
  \bibinfo{journal}{{IEEE} Commun. Surveys Tuts.}
  \textbf{\bibinfo{volume}{21}}, \bibinfo{pages}{970} (\bibinfo{year}{2019}),
  ISSN \bibinfo{issn}{1553-877X}.

\bibitem[{\citenamefont{Ryan and Lin}(2009)}]{RL09}
\bibinfo{author}{\bibfnamefont{W.}~\bibnamefont{Ryan}} \bibnamefont{and}
  \bibinfo{author}{\bibfnamefont{S.}~\bibnamefont{Lin}},
  \emph{\bibinfo{title}{Channel codes -- {Classical} and modern}}
  (\bibinfo{publisher}{Cambridge University Press}, \bibinfo{address}{New York,
  NY, USA}, \bibinfo{year}{2009}).

\bibitem[{\citenamefont{Cover and Thomas}(2006)}]{CovTho:06}
\bibinfo{author}{\bibfnamefont{T.~M.} \bibnamefont{Cover}} \bibnamefont{and}
  \bibinfo{author}{\bibfnamefont{J.~A.} \bibnamefont{Thomas}},
  \emph{\bibinfo{title}{Elements of Information Theory}}
  (\bibinfo{publisher}{John Wiley \& Sons, Inc.}, \bibinfo{address}{Hoboken,
  NJ}, \bibinfo{year}{2006}), \bibinfo{edition}{2nd} ed.

\end{thebibliography}
